\documentclass[12pt,preprint]{aastex}

\usepackage{graphicx,epsfig,natbib,color}
\usepackage{lscape}
\usepackage{graphicx}
\usepackage{epsfig}
\usepackage{natbib}

\begin{document}
\title{Observing Flux Rope Formation During the Impulsive Phase of a Solar Eruption}

\author{X. Cheng\altaffilmark{1,2,3}, J. Zhang\altaffilmark{2}, Y. Liu\altaffilmark{4}, M. D. Ding\altaffilmark{1,3}}

\affil{$^1$ Department of Astronomy, Nanjing University, Nanjing 210093, China}
\affil{$^2$ School of Physics, Astronomy and Computational Sciences, George Mason University, 4400 University Drive, MSN 6A2, Fairfax, VA 22030, USA}\email{jzhang7@gmu.edu}
\affil{$^3$ Key Laboratory for Modern Astronomy and Astrophysics (Nanjing University), Ministry of Education, Nanjing 210093, China}\email{dmd@nju.edu.cn}
\affil{$^4$ W.W. Hansen Experimental Physics Laboratory, Stanford University, Stanford, CA 94305, USA}

\begin{abstract}
Magnetic flux rope is believed to be an important structural component of coronal mass ejections (CMEs). While there exist much observational evidence of the flux rope after the eruption, e.g., as seen in remote-sensing coronagraph images or in-situ solar wind data, the direct observation of flux ropes during CME impulsive phase has been rare. In this Letter, we present an unambiguous observation of a flux rope still in the formation phase in the low corona. The CME of interest occurred above the east limb on 2010 November 03 with footpoints partially blocked. The flux rope was seen as a bright blob of hot plasma in AIA 131 \AA\ passband (peak temperature $\sim$11 MK) rising from the core of the source active region, rapidly moving outward and stretching upward the surrounding background magnetic field. The stretched magnetic field seemed to curve-in behind the core, similar to the classical magnetic reconnection scenario in eruptive flares. On the other hand, the flux rope appeared as a dark cavity in AIA 211 \AA\ passpand (2.0 MK) and 171 \AA\ passband (0.6 MK); in these relatively cool temperature bands, a bright rim clearly enclosed the dark cavity. The bright rim likely represents the pile-up of the surrounding coronal plasma compressed by the expanding flux rope. The composite structure seen in AIA multiple temperature bands is very similar to that in the corresponding coronagraph images, which consists of a bright leading edge and a dark cavity, commonly believed to be a flux rope.

\end{abstract}
Online-only material: color figures, animations

\section{Introduction}
Coronal mass ejections (CMEs) are a large-scale solar activity, releasing a vast amount of plasma and magnetic field into the interplanetary space and may cause severe disturbances of the space environment \citep{gosling93,webb94}. In the classical eruptive flare model \citep{carmichael64,sturrock66,hirayama74,kopp76}, magnetic reconnection is regarded as the fundamental energy releasing mechanism. In this kind of models, the originally anchored magnetic field lines are stretched upward by an erupting structure, and the field lines underneath are pushed toward the center forming a current sheet (CS). The reconnection is induced in the CS, creating new closed field lines (as seen in a 2-D projection) at both tips of the CS. Magnetic reconnection is also believed to produce energetic particles. These particles stream down along the newly reconnected field lines and produce the well observed flare ribbons and post-flare loops, as well as enhancing soft X-ray and hard X-ray emissions \citep{forbes96,priest02}. Above the reconnection region, poloidal magnetic fluxes are injected into the erupting structure, forming and/or strengthening a flux rope. The enhanced poloidal flux in a flux rope would increase the upward Lorentz self-force and thus further accelerate the flux rope outward forming a CME \citep{chen96}. From observations, \citet{zhang01,zhang04} found that a flare-associated CME usually undergoes three distinct phases of evolution: the slow-rising initiation phase, the impulsive acceleration phase, and the near-constant propagation phase. The impulsive acceleration phase of the CME coincides in time very well with the rise phase of the flare \citep[also see,][]{bur04,temmer08,cheng10a}. This temporal coincidence strongly implies that the same solar eruption could produce both the flare and CME phenomena, which are coupled together through magnetic reconnection.

The flux rope, a set of magnetic field lines winding around an axis, is a key feature in classical eruptive flare models and many CME observations. Helical magnetic field structures, the so-called magnetic clouds, have been frequently found in in-situ solar wind observations \citep{burlaga82,klien82}. The internal helical structure of CMEs can be directly observed by coronagraphs \citep[e.g.,][]{dere99}. Moreover, coronagraph images show that a CME usually consists of a three-part structure: the bright front or leading edge (LE), the enclosed dark cavity and the inner bright core \citep{illing83}. The existence of a flux rope could explain the dark cavity structure of a CME \citep[e.g.,][]{gibson06,wang10}. Flux ropes might be preceded by sigmoid structures in the corona before the eruption \citep{canfield99,mckenzie08}. Recent observations showed that the sigmoid structures, which originally consisted of two opposite J-shaped loops, could be transformed into a flux rope \citep{liu10,green11}. \citet{torok03,torok05} and \citet{kliem06} studied the instability of the flux rope and proposed that the kink and/or torus instability can trigger the CME. \citet{olmedo10} made a more detailed study and proposed the partial torus instability of the flux rope. Through nonlinear force-free field models, several authors were able to reconstruct a flux rope configuration in the corona from observed photospheric magnetic fields \citep[e.g.,][]{canou09,guo10,cheng10b}. Considerable efforts have also been made in numerical MHD simulations to study the formation and dynamic behavior of flux ropes \citep{amari00, amari03, aulanier10,fan03,fan04}.

While it is widely accepted that the flux rope is either formed during the CME eruption or exists prior to the eruption, the direct observation of the flux rope structure during the eruption process has been rare \citep[a plausible case was reported by][]{gary04}. In this Letter, we report the observation of a CME event by \textit{Solar Dynamic Observatory} (\textit{SDO}), \textit{Solar Terrestrial Relations Observatory} (\textit{STEREO}), and \textit{Solar and Heliospheric Observatory} (\textit{SOHO}), which clearly showed a flux rope structure during the impulsive acceleration phase of the CME. The observed structure is also found to be consistent with the classical eruptive flare model described earlier. In Section 2, we present the observations and results, which are followed by a summary and discussions in Section 3.

\section{Observations and Results}

\subsection{Instruments}

The Atmospheric Imaging Assembly (AIA) on board \textit{SDO} (the first observatory of NASA's Living With a Star Program) is designed to image the multi-layered solar atmosphere, including photosphere, chromosphere, transition region, and corona in ten narrow UV and EUV passbands with unprecedented cadence (12 seconds), high spatial resolution (0.6$''$ pixel size) and large field of view (1.3$R_\odot$). AIA's capacity of high cadence and multiple temperature provides the opportunity of observing the formation of the flux rope in the low corona during the impulsive acceleration phase of a CME. In addition, the Sun Earth Connection Coronal and Heliospheric Investigation (SECCHI) on board \textit{STEREO} provides the direct observations of CMEs in the coronagraphs. The Large Angle Spectroscopic Coronagraph (LASCO) on board \textit{SOHO} provides the CME images from one additional perspective.

\subsection{Formation of Flux Rope}

\begin{figure} 
     \vspace{-0.0\textwidth}    
     \centerline{\hspace*{0.02\textwidth}
               \includegraphics[width=0.9\textwidth,clip=]{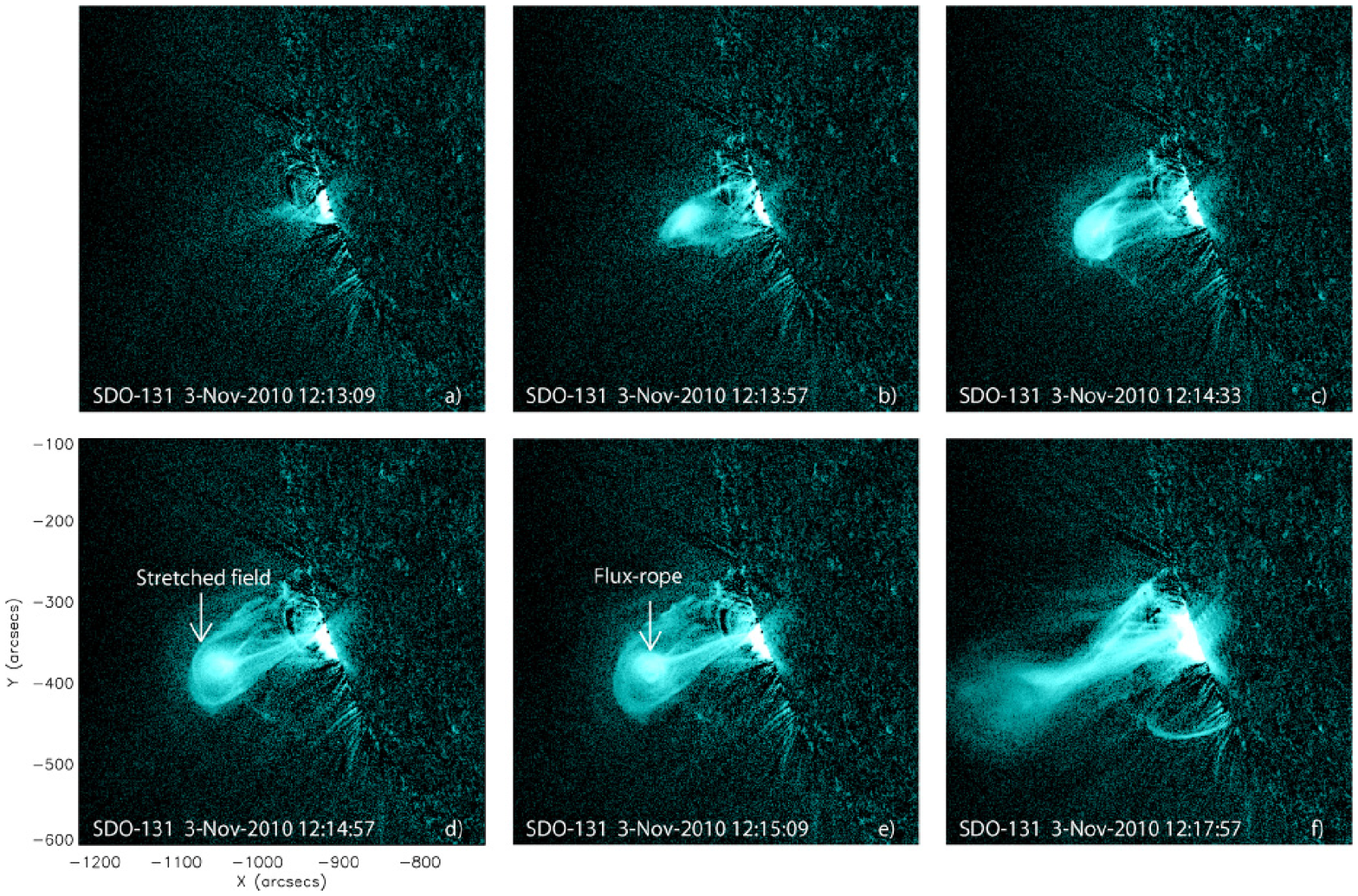}
               }

\vspace{0.0\textwidth}   
\caption{AIA 131 {\AA} ($\sim$11 MK) base-difference images of the solar eruption on 2010 November 03. The base image is the one at 12:00 UT. Stretched overlying magnetic field lines and flux rope are indicated in panel d and e, respectively.}
\label{f1}
      (An animation is available at http://spaceweather.gmu.edu/xcheng/incoming/1103/.)

\end{figure}

On 2010 November 3, a C4.9 class soft X-ray flare occurred at the east limb of the Sun, which started at $\sim$12:07 UT and peaked at $\sim$12:21 UT. With \textit{STEREO}-B coronal images, which showed the flare on the disk, we find that the flare was located at $\sim$S20E96 from the perspective of the Earth, and thus a fraction of the soft X-ray emission from the flare was occulted. Through inspecting LASCO/C2 images, it is confirmed that the flare was associated with a limb CME. In AIA 131 \AA\ images, a blob of plasma first appeared above the solar limb at $\sim$12:13 UT (Figure \ref{f1}(a)--(c)). The circular blob was only visible as enhanced emission in the 94 {\AA} and 131 {\AA} images, but not in any other AIA passbands, which indicate that the plasma in the blob was very hot with temperatures as high as $\sim$7 to 11 MK \citep[see][for response functions of AIA channels]{odwyer10}. At $\sim$12:14 UT, the blob of hot plasma started to push its overlying magnetic field upward and form a sharp edge (Figure \ref{f1}(d)). The overlying field lines seem to be stretched up continuously. Immediately below the blob, there appeared a Y-type magnetic configuration, in which a bright thin line extended downward.

This kind of structure is largely consistent with the classical eruptive flare model \citep{carmichael64,sturrock66,hirayama74,kopp76,shibata95,lin00}, in which a CS forms below the rising flux rope, and magnetic reconnection induced in the CS converts the stretched surrounding magnetic field into the new poloidal flux of the flux rope. The observed plasma blob is likely to be the flux rope as proposed in the classical theory (more evidence will be given later). The observations may indicate that the flux rope was undergoing the forming and growing phase (Figure \ref{f1}(b)--(f)). If this is the case, it might be the first time that the formation of flux rope is directly observed in the low corona. The flux rope can be seen, not only because of the superb AIA imaging capacity, but also the favorable orientation of the flux rope; it is likely that the flux rope was oriented edge on, i.e., the axis of the flux rope lied more or less along the line of sight. Nevertheless, $STEREO$-B EUVI 195~\AA\ images did not reveal a definite orientation of the eruption structure, since a post-eruption loop arcade did not show up for this event in which the flare had a short duration in soft X-ray.

\begin{figure} 
     \vspace{-0.0\textwidth}    
     \centerline{\hspace*{0.02\textwidth}
               \includegraphics[width=0.9\textwidth,clip=]{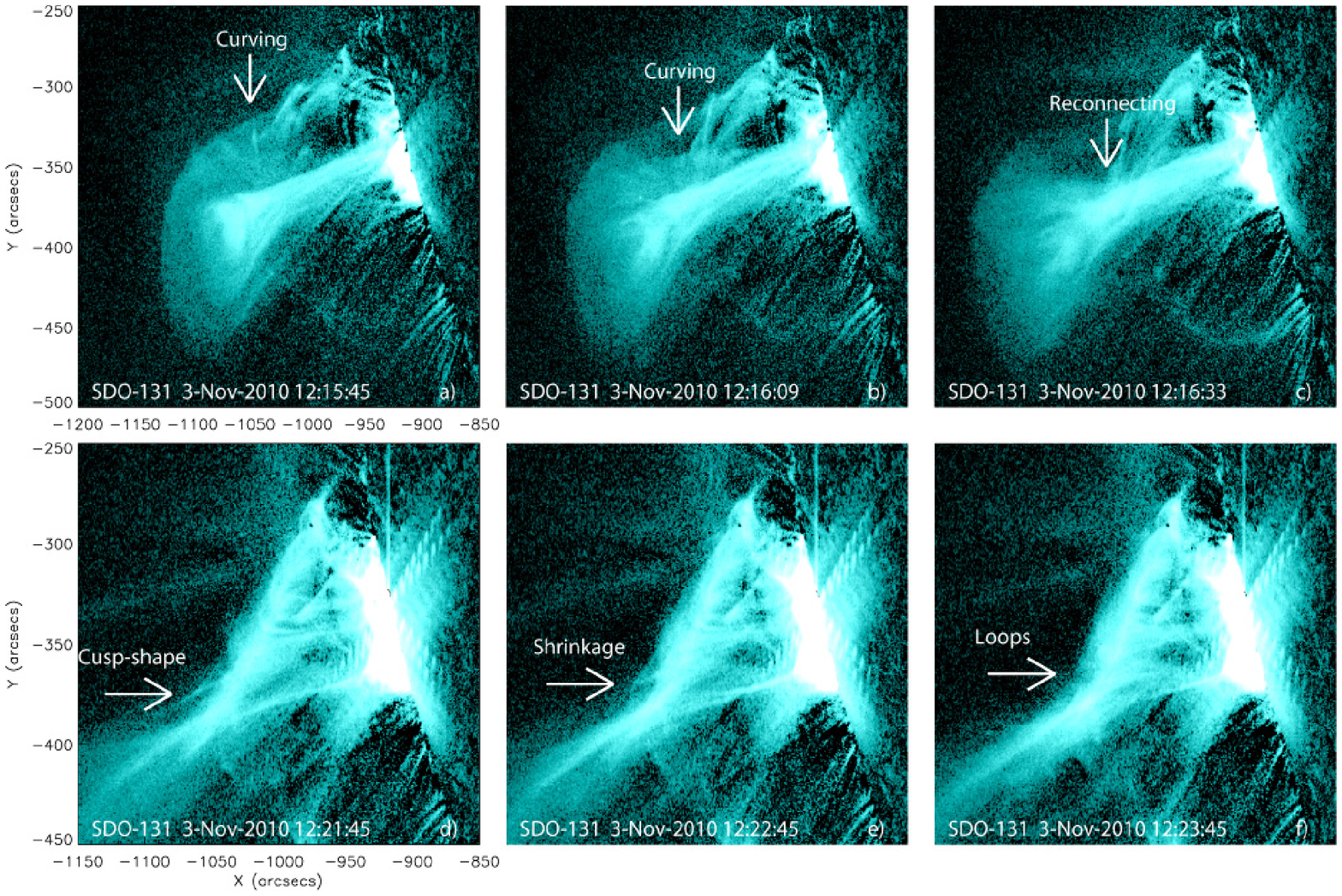}
                              }
     \centerline{\hspace*{0.065\textwidth}
               \includegraphics[width=0.855\textwidth,clip=]{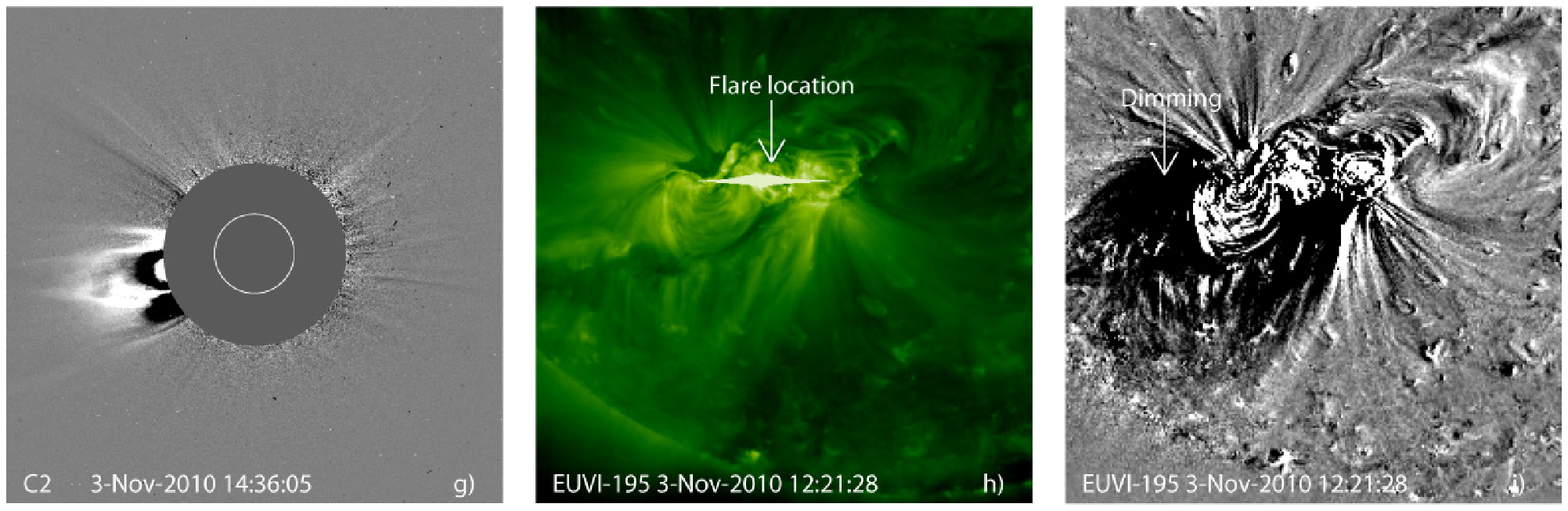}
}
     \vspace{0.0\textwidth}   
\caption{(a--f) Same as in Figure 1 but with a smaller FOV. The curving of the stretched overlying magnetic field lines toward the dissipation region is indicated in panel a--c. The shrinkage process of the newly reconnected magnetic field lines at the low tip of the CS is illustrated in panel d--f. (g) Running difference image of the CME from LASCO/C2 observations. (h) EUVI 195 {\AA} image showing the flaring source region. (i) Running difference image of the eruption source region showing the corresponding dimming.}
   \label{f2}

 \end{figure}

The newly added poloidal flux will increase the outward Lorentz self-force of the flux rope, resulting in faster acceleration of the structure. In response to this outflow of magnetic flux, the inflow speed of the anti-parallel ambient magnetic field toward the dissipating CS is expected to be strengthened, resulting in a faster magnetic reconnection rate, which further increases the poloidal flux injection rate. Thus, this is a positive feedback process between magnetic reconnection and flux rope formation/escaping. The observations are consistent with this kind of process. As the flux rope continually rose, the overlying fields seemed to be stretched as shown in Figure \ref{f2}(a)--(c). Further, it is seen that the stretched magnetic field lines curved toward the dissipation region at 12:15:45 UT (Figure \ref{f2}(a)). The curved magnetic field lines may arrive at the reconnection region and then participate in the reconnection (12:16:33 UT, Figure \ref{f2}(c)). On the other hand, we note that the curving process only lasted for $\sim$2 minutes. The curving of other field lines might be invisible own to a higher temperature ($>$11MK). It is also worth noting that this very hot plasma present in the inflow of the region is difficult to be explained by the classical eruptive model. There might be heating taking place in the region before the reconnection.

In addition to the inflow from two sides of the CS, we also observe the shrinkage of magnetic field lines underneath the CS, indicating the ongoing process of magnetic reconnection. At the lower Y-type point of the CS, the reconnected field lines were originally cusp-shaped. However, thanks to the magnetic tension force, they then shrank and became loop-shaped \citep{forbes96,priest02}. After the CS forming, probably as early as at 12:14:45, the reconnection in the CS formed the post-flare loops under the CS. As the reconnection was ongoing, the Y-type and reversed Y-type configuration appeared and they was connected by a short and thin feature (e.g., Figure \ref{f1}(f)). In order to illustrate  the shrinkage, we plot the reversed Y-type structure imaged in 131 {\AA} in Figure \ref{f2}(d)--(f). The cusp-shaped field lines appeared near the crossing of the reversed Y-type configuration (Figure \ref{f2}(d)); they may be just formed by the reconnection of the two antiparallel legs of the stretched field lines. Subsequently, the field line with the cusp-shape rapidly shrank (Figure \ref{f2}(e)) and then changed to the semi-circular post-flare loops (Figure \ref{f2}(f)). The details of the shrinkage process can be better seen in the attached movie.

\begin{figure} 
     \vspace{-0.0\textwidth}    
\centerline{\hspace*{0.0\textwidth}
               \includegraphics[width=0.48\textwidth,clip=1]{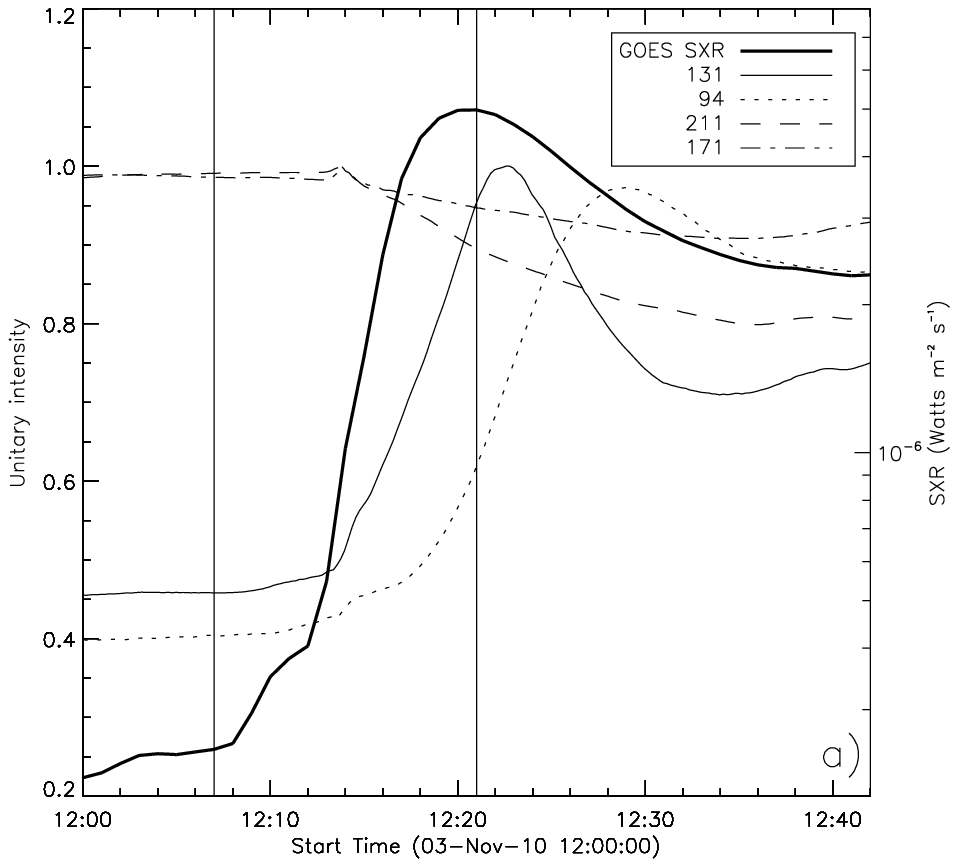}
               \includegraphics[width=0.48\textwidth,clip=]{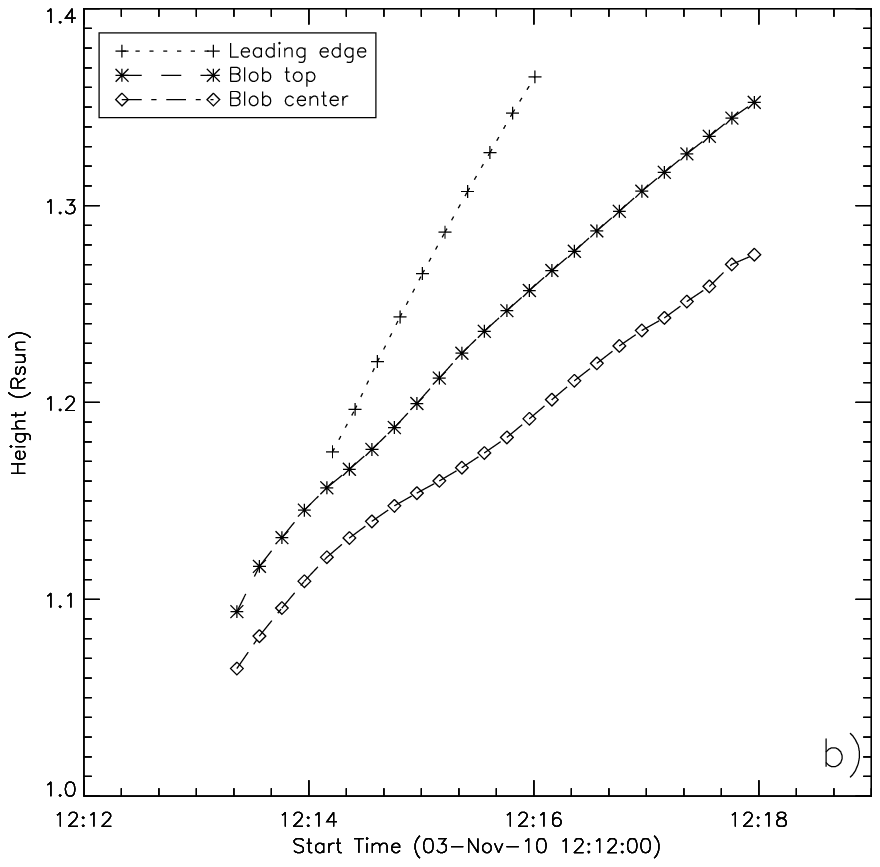}
            }

     \vspace{0.0\textwidth}   
     
\caption{(a) $GOES$ soft X-ray 1--8 {\AA} flux and 131 {\AA} ($\sim$11 MK), 94 {\AA} ($\sim$7 MK), 211 {\AA} ($\sim$2 MK), and 171 {\AA} ($\sim$0.6 MK) integral intensity in the FOV of Figure 1, which are normalized by their maxima. Two vertical lines show the onset and peak times of the flare in $GOES$ X-ray. (b) The height-time profiles of the LE, the blob top, and blob center in the impulsive phase.}

   \label{f3}

   \end{figure}

In order to compare the emission produced in different AIA passbands, we plot the temporal profiles of the intensity integrated over the eruption region (the region shown in Figure~\ref{f1}) for 131 {\AA}, 94 {\AA}, 211 {\AA}, and 171 {\AA} respectively in Figure \ref{f3}, along with the $GOES$ soft X-ray 1--8~\AA\ profile. It is found that the profile of the 131 {\AA} emission was similar to that of the soft X-ray, but with a systematic delay of $\sim$2 minutes. The emission profile of 94 {\AA} was also similar to that of 131 {\AA} but with a delay of $\sim$7 minutes. The delay is probably caused by the cooling from high to low temperature of the plasma located in and surrounding the reconnection region \citep{aschwanden01,warren03}. On the other hand, the integrated intensity of 211 {\AA} and 171 {\AA} started to decrease at 12:10 UT, probably caused by the reconnection heating moving the plasma out of the temperature range sensitive to these passbands.

\subsection{Multi-component and Multi-temperature Structure}

In this section, we discuss the multi-component and multi-temperature nature of the CME structure in the CME formation phase or impulsive phase. In addition to the flux rope, the formation of CME LE was also observed. It appeared that the top of the accelerated flux rope was compressing the surrounding plasma and producing the compression front with enhanced plasma density. The LE was formed from the successive stacking of the compression front as the flux rope was expanding and rising, but not the LE does not represent the real rising of the same stretched field lines. Therefore, the LE velocity of the CME is the apparent velocity of the compression wave running ahead of the CME flux rope, while the flux rope acts as the driver of the compression wave. By tracking the LE, the blob top, and the blob center, we plot their height-time profiles in Figure~\ref{f3} (b), from which one can find that the average velocities of the LE, blob top and blob center in the FOV of AIA are $\sim$1200 km/s, 630 km/s, 500 km/s, respectively. Both the LE and cavity later appeared in the FOV of LASCO/C2 (Figure \ref{f2}(g)).

\begin{figure} 
     \vspace{-0.0\textwidth}    
     \centerline{\hspace*{0.02\textwidth}
               \includegraphics[width=0.9\textwidth,clip=]{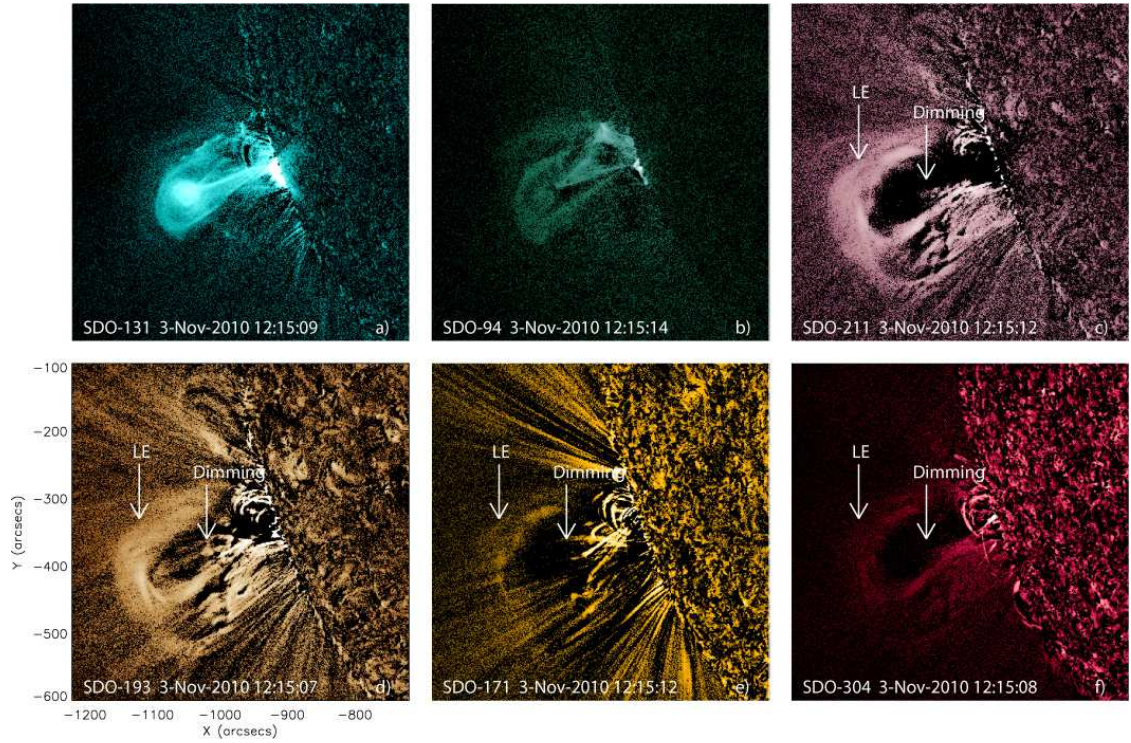}
               }

     \vspace{0.0\textwidth}   

\caption{AIA base-difference images of the solar eruption on 2010 November 03 at six passbands (131 {\AA} ($\sim$11 MK), 94 {\AA} ($\sim$7 MK), 211 {\AA} ($\sim$2 MK), 193 {\AA} ($\sim$1 MK), 171 {\AA} ($\sim$0.6 MK), and 304 {\AA} ($\sim$0.05 MK)). All images are at $\sim$12:15 UT subtracting the corresponding base images at $\sim$12:00 UT. Leading edge and dimming features are indicated by the arrows.}
   \label{f4}
         (An animation is available at http://spaceweather.gmu.edu/xcheng/incoming/1103/.)
   \end{figure}

Another important finding from the observations is that the CME has multiple structural components well organized by the temperature. The appearance of the eruption at $\sim$12:15 UT is shown in six different passbands (131 {\AA}, 94 {\AA}, 211 {\AA}, 193 {\AA}, 171 {\AA}, and 304 {\AA}) in Figure \ref{f4}. Note that the images are difference images by subtracting the corresponding base images at $\sim$12:00 UT. It is evident that the center part of the CME is of high temperature (as high as $\sim$7 to 11 MK) (Figure \ref{f4}(a) and (b)) \citep[see also][]{reeves11}. The temperature of the LE may distribute from $\sim$0.05 to 2.5 MK, since the LE is visible at 211 {\AA}, 193 {\AA}, 171 {\AA}, and 304 {\AA} (Figure \ref{f4}(c)--(f)). From the composite images with three passbands (171 {\AA}, 211 {\AA}, and 131 {\AA}) (attached movie), one can clearly identify that the CME has a multi-temperature structure (a hot core and a cool LE).

The multi-temperature structure of the CME also sheds light on the origin of the dimming region in the impulsive phase. It is usually believed that the dimming is caused by the loss of mass in the low corona \citep[e.g.,][]{thompson98}. However, the multi-temperature observations suggest that the high temperature in the center part of the CME also play a role in producing the observed dimming. Hot plasma induces the emission at higher temperature and is less visible or invisible in the passbands of lower temperature. As shown in Figure \ref{f4}, the shape of the dimming region (at 211 {\AA} and 171 {\AA}) was almost consistent with the shape of the CME hot core (at 131 {\AA}). It reveals that the dimming may result from the absence of the low temperature emission, which holds at least during the early formation phase of the CME. Note that, the 193 {\AA} also includes partial emission of the hot plasma so that the dimming present at 193 {\AA} was blended by a few hot emission from the CME center part (also see the 195 {\AA} difference image in Figure \ref{f2}(i)).

\section{Summary and Discussions}

\begin{figure} 
     \vspace{-0.0\textwidth}    
     \centerline{\hspace*{0.0\textwidth}
               \includegraphics[width=0.7\textwidth,clip=]{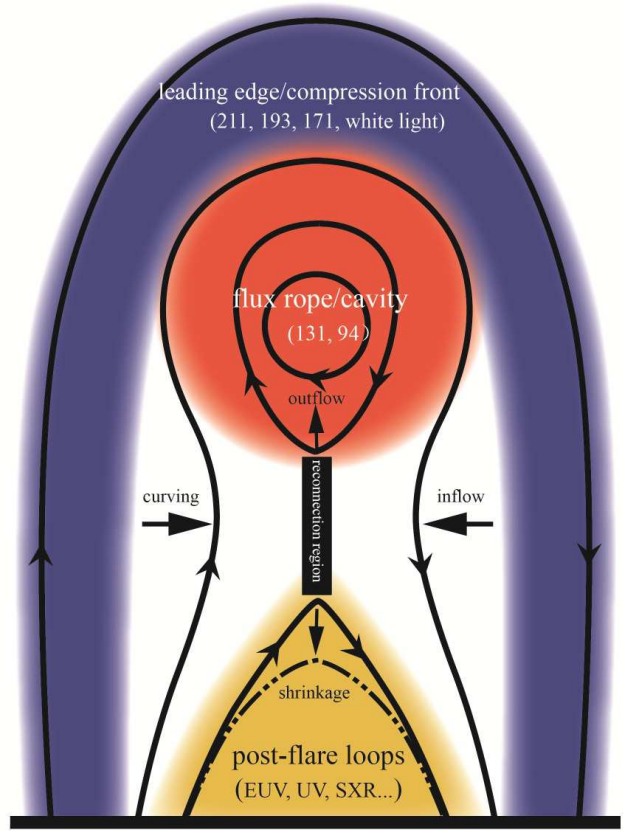}
              }
     \vspace{0.0\textwidth}   

\caption{Schematic drawing of the multi-component multi-temperature of the solar eruption in the low corona. Red denotes the hot flux rope/cavity; blue for the cool LE/compression front; yellow for reconnected post-flare loops with mixed temperatures.}
   \label{f5}

   \end{figure}

We summarize our observations of the formation of the flux rope and the multi-temperature structure of the CME in a schematic model shown in Figure~\ref{f5}. The black lines denote the magnetic field configuration. The different features relevant to the magnetic reconnection, including the flux rope, LE, curving, and shrinkage, are indicated by the arrows. The multi-temperature components of the CME structure are shown in different background colors; the texts indicate the effective observing passbands for different features.

It is most likely that the flux rope structure was forming in the impulsive acceleration phase of the CME. The formation process was highly consistent with the classical eruptive flare model \citep{carmichael64,sturrock66,hirayama74,kopp76}. The ejection of a blob of hot plasma, which is the possible pre-flux-rope structure, moved upward and stretched the overlying restraining magnetic field lines. Underneath the blob, the stretched field lines were pushed toward the line-shaped CS and reconnected in the CS. For this particular event, it is largely the magnetic reconnection that produced the bulk of the flux rope structure through injecting the poloidal magnetic flux. The ejection of the magnetic flux rope caused a decrease of magnetic pressure at both sides of the CS and thus led to an inflow toward the CS \citep{yokoyama01}. The inflow curved the legs of the newly stretched field lines toward the reconnection region and made the reconnection continuing \citep{zhang06,cheng10a,lin00}. Under the lower tip of the CS, the reconnected magnetic field lines has initially a cusp shape and then shrank to become the semi-circular post-flare loops \citep{svestka87,forbes96}. In the upper tip of the CS, the newly formed poloidal magnetic field lines were added into the flux rope, carrying with them the heat generated in the reconnection region. Therefore, the temperature in the center part of the CME was higher than the surrounding part. This resulted in the multi-temperature structure of the CME.

It is worthy noting that the counterpart of the flux rope feature we are reporting here may have been noticed in earlier observations, e.g, the plasmoid above soft X-ray loops \citep{shibata95,ohyama98,reeves11}. \citet{ohyama98} derived the temperature of the plasmoid at $\sim$10 MK, which was similar to the response temperature of 131 {\AA}. Furthermore, we believe that the blob of the hot plasma, which later formed the full-fledged flux rope through magnetic reconnection, was originated from either a sheared core field in the low corona \citep{moore92,antiochos99} or a weakly twisted flux rope existing prior to the eruption \citep{chen96,torok05,kliem06,cheng10b,guo10,olmedo10,liu10}.


\acknowledgements
We thank the anonymous referee for his/her valuable comments that helped to improve the paper. SDO is a mission of NASA's Living With a Star Program. X.C., and M.D.D. are supported by NSFC under grants 10673004, 10828306, and 10933003 and NKBRSF under grant 2011CB811402. X.C. is also supported by the scholarship granted by the China Scholarship Council (CSC) under file No. 2010619071. J.Z., is supported by NSF grant ATM-0748003 and NASA grant NNG05GG19G.


\bibliographystyle{apj}




\end{document}